\documentstyle[preprint,aps,epsf]{revtex}
\begin{document}
\def\be{\begin{equation}}
\def\ee{\end{equation}}
\def\bea{\begin{eqnarray}}
\def\eea{\end{eqnarray}}
\def\beas{\begin{eqnarray*}}
\def\eeas{\end{eqnarray*}}
\def\no{\nonumber}
\def\a{\alpha}
\def\b{\beta}
\def\g{\gamma}
\def\G{\Gamma}
\def\d{\delta}
\def\D{\Delta}
\def\e{\epsilon}
\def\ve{\varepsilon}
\def\k{\kappa}
\def\m{\mu}
\def\n{\nu}
\def\L{\Lambda}
\def\l{\lambda}
\def\mn{\mu\nu}
\def\mnl{\mu\nu\lambda}
\def\O{\Omega}
\def\p{\phi}
\def\vp{\varphi}
\def\P{\Phi}
\def\ps{\psi}
\def\r{\rho}
\def\s{\sigma}
\def\th{\theta}
\def\bp{{\bf p}}
\def\bA{{\bf A}}
\def\br{{\bf r}}
\def\t{\tilde}
\def\bra{\langle}
\def\ket{\rangle}
\newcommand{\lbra}{\left\langle}
\newcommand{\rket}{\right\rangle}
\def\cd{\cdot}
\def\eq{\equiv}
\def\i{\infty}
\def\pa{\partial}
\def\ra{\rightarrow}
\def\lra{\longrightarrow}
\def\ri{r\rightarrow\infty}
\def\inttd{\int\! d^2\!\vvr\,}
\def\inttdp{\int\! d^2\!\vvr'\,}
\def\intdx{\int\! d^4\!x\,}
\newcommand{\np}[1]{Nucl.\ Phys.\ {\bf {#1}}}
\newcommand{\plt}[1]{Phys.\ Lett.\ {\bf {#1}}}
\newcommand{\mpl}[1]{Mod.\ Phys.\ Lett.\ {\bf {#1}}}
\newcommand{\pr}[1]{Phys.\ Rev.\ {\bf {#1}}}
\newcommand{\prlt}[1]{Phys.\ Rev.\ Lett.\ {\bf {#1}}}
\newcommand{\ijmp}[1]{Int.\ J.\ Mod.\ Phys.\ {\bf {#1}}}
\newcommand{\rmps}[1]{Rev.\ Mod.\ Phys.\ {\bf {#1}}}
\newcommand{\prp}[1]{Phys.\ Rep.\ {\bf {#1}}}
\newcommand{\anp}[1]{Ann.\ Phys. (N. Y.)\ {\bf {#1}}}
\newcommand{\cmp}[1]{Comm.\ Math.\ Phys.\ {\bf {#1}}}
\newcommand{\jmp}[1]{J.\ Math.\ Phys.\ {\bf {#1}}}
\newcommand{\jp}[1]{J.\ Phys.\ {\bf {#1}}}
\newcommand{\spj}[1]{Sov.\ Phys.\ JETP {\bf {#1}}}
\newcommand{\sjnp}[1]{Sov.\ J.\ Nucl.\ Phys.\ {\bf {#1}}}
\newcommand{\ptp}[1]{Prog.\ Theor.\ Phys.\ {\bf {#1}}}
\def\erf{\mbox{Erf\,}}
\preprint{hep-th/9706184\hspace{-28mm}\raisebox{2.4ex}{SNUTP 97-085}}
\title{Second Virial Coefficient of Anyons without Hard Core}
\author{Chanju Kim\thanks{cjkim@ctp.snu.ac.kr}}
\address{Center for Theoretical Physics\\
         Seoul National University\\
         Seoul, 151-742, Korea}
\maketitle
\begin{abstract}
We calculate the second virial coefficient of anyons whose wave function
does not vanish at coincidence points. This kind of anyons appear naturally
when one generalizes the hard-core boundary condition 
according to self-adjoint extension method in quantum mechanics, 
and also when anyons are treated field theoretically by applying
renormalization procedure to nonrelativistic Chern-Simons field theory.
For the anyons which do not satisfy hard-core boundary condition, it is
argued that the other scale-invariant limit is more relevant in 
high-temperature limit where virial expansion is useful. 
Furthermore, the cusp existing at the bosonic point for hard-core anyons
disappears in general; 
instead it is shown that a new cusp is generated at
the fermionic point. A physical explanation is given.
\end{abstract}
\vspace{10mm}
\newpage
In this letter we would like to calculate the second virial coefficient of a
system of anyons \cite{anyon} obtained by generalizing the usual hard-core 
boundary condition at coincidence points 
according to the self-adjoint extension method \cite{SA}, and discuss its
interesting physical properties.
To motivate our calculation, let us briefly summarize the present status.

Anyons are particles exhibiting fractional statistics characterized by
a statistical parameter $\a$ which interpolates between bosonic and fermionic
cases. Quantum mechanically, they can be regarded as flux-charge composites
and their interaction is essentially Aharonov-Bohm type \cite{AB}. 
From the field
theoretic point of view, they can be obtained if the Schr\"odinger field is
coupled to a Chern-Simons gauge field \cite{hagen,JP}. 
The connection of these two
descriptions is now well established and an interesting subject here is the
relation between the method of self-adjoint extension in quantum mechanics and
the regularization-renormalization procedure in field theory where 
$|\p|^4$-type contact interaction is necessarily 
introduced \cite{BL,AM,AD,KL}. Recently, an all-order analysis
has been carried out to establish the precise correspondence and the role of
contact term has been emphasized in ref.\ \cite{KL}.

In the literature, regularity of wave functions is often assumed and 
it leads to the hard-core boundary condition \cite{anyon}. 
Although this hard-core case
is conceptually simple and easier to treat, there is no {\it a priori} 
reason to assume such a boundary condition in general; a real system under
study should determine the relevant boundary condition eventually.
If we relax the regularity requirement, allowing wave functions 
to diverge at coincidence points 
according to the method of self-adjoint extension, 
we get a one-parameter family of boundary conditions 
which in general introduces a scale in the theory. (The corresponding 
anyons were called ``colliding anyons" \cite{BoS}.) 
The scale invariance of the theory is 
restored only in the limit that the scale parameter goes to either 
zero or infinity. Field theoretically, when
the strength of the $|\p|^4$-interaction is equal to the critical value
$\l = \pm\frac{2\pi|\a|}{m}\equiv \pm \l_0$, the field theory becomes
ultraviolet finite and scale invariant \cite{KL}. 
These critical values have some more
interesting properties. The repulsive case $\l =\l_0$ corresponds to the
usual hard-core boundary condition, while in the attractive case $\l = -\l_0$,
it is known that the classical theory admits static soliton solutions
satisfying a self-dual equation \cite{JP}. Also, for $\l \rightarrow -\l_0$, 
after performing renormalization, one can relate the renormalized coupling
$\l_{\text{ren}}$ to a nonzero finite value of the self-adjoint extension 
parameter in a specific way \cite{BB,KL}. 

Since the free anyon gas has not been exactly solved, one usually resorts to
the virial expansion to study thermodynamic properties in high-temperature
low-density limit. For example, the second virial coefficient was calculated
for anyons with hard core in ref.\ \cite{arovas,CGO,LF,HK}. 
Here an interesting fact 
is that, according to recent analyses \cite{BB,KL}, $\l_{\text{ren}} = \l_0$ 
(hard-core case) is an infrared fixed point, 
while $\l_{\text{ren}} = -\l_0$ is an ultraviolet 
fixed point, with $\l_{\text{ren}}$ flowing from $\l_0$ to $-\l_0$ as the
renormalization scale increases. This implies that the virial 
expansion for usual anyons with hard core is not much relevant for the 
case of colliding anyons in high-temperature limit where the 
virial expansion is
useful. Therefore it may be interesting to calculate the second virial 
coefficient for anyons with an arbitrary boundary condition and study the 
behavior as the scale changes. Indeed, we will see that 
the result
of hard-core case does not represent a typical behavior of the other cases. 
For example, the
cusp existing at bosonic point $\a=0$ in hard-core case is a very special
feature of scale-invariant limit which does not occur in general; 
instead a new cusp is
generated at the {\it fermionic} point $|\a|=1$ except for the hard core case. 
We will give a physical interpretation of this result later.

Here we should mention earlier works \cite{moroz,GMS} which have a partial 
overlap with this paper. They calculated the second virial coefficient 
for the case that the system has a bound state. However they did not 
pay much attention to physical properties for general boundary conditions
except at the fixed point, $\l_{\text{ren}}=-\l_0$. Also, we believe that
our method is simpler and physically more transparent.

Let us start with the Hamiltonian of a system of free anyons with mass $m$ and
statistical parameter $\a$ in bosonic description
\be
H=\sum_n\frac1{2m}(\bp_n-\a\bA_n)^2\,,
\ee
where
\be
A_n^i\equiv \e^{ij}\sum_{m(\neq n)} \frac{x_n^j - x_m^j}{|\br_n - \br_m|^2}
\ee
is the Aharonov-Bohm-type vector potential and $\br = (x_n^1, x_n^2)$ denotes
the position of the $n$-th particle. As mentioned above, 
this Hamiltonian does not specify the system completely and 
one has to give a
suitable boundary conditions when particle positions coincide; we will give 
the form later.

In two dimensions, the second virial coefficient $B(T)$ is given by
\cite{huang}
\be  \label{bt}
B(T) = A(\frac12 - Z_2/Z_1^2)\,,
\ee
where $A$ is the area of the system and $Z_1$ and $Z_2$ are the one-body and
two-body partition functions, respectively. They are calculated as
\bea \label{4}
Z_1 &=& A \l_T^{_2} \,,\no\\
Z_2 &=& 2A\l_T^2 Z_{\text{rel}}\,,
\eea
where $\l_T = (2\pi/mk_BT)^{1/2}$ is the thermal wavelength and
$Z_{\text{rel}}= {\rm Tr} e^{-\b H_{\text{rel}}}$ is the single partition 
function of the relative dynamics with
\bea
H_{\text{rel}}&=&\frac1m (\bp-\a \bA)^2, \no\\
A^{i}&\equiv&\frac{\e^{ij}x^j}{r^2}, \quad (r\equiv |{\bf r}|).
\eea

After the separation of variables $\ps(\br) = e^{in\th}R_n(r)$, the
Schr\"odinger equation for $H_{\text{rel}}$ reads
\be  \label{bessel}
\frac1m\left[-\frac1r \frac{d}{dr}r\frac{d}{dr} + \frac{(n+\a)^2}{r^2}\right]
R_n(r) = ER_n(r) \equiv \frac{k^2}{m}R_n(r)\,,
\ee
where $n$ is even since we are working in bosonic description. 
Also, since $\a$
enters (\ref{bessel}) only through the combination $n+\a$, we may restrict
$\a$ to the interval $[-1,1]$ without loss of generality in considering
the energy spectrum \cite{anyon}. 
Equation (\ref{bessel}) is nothing but the Bessel equation and
the general solution is of the form
\be \label{solution}
R_n(r) = AJ_{|n+\a|}(kr) + BJ_{-|n+\a|}(kr)\,,
\ee
where $A$, $B$ are constants. For $n\neq 0$ the constant $B$ must be chosen to
be zero for square integrability, but for $n=0$ ($s$-wave case)
arbitrary $A$ and $B$ are allowed. This leads to
an above-mentioned one-parameter family of boundary conditions for 
$s$-wave \cite{MT},
\be \label{bc}
\lim_{r\ra 0}\left\{r^{|\a|}R_0(r)-\frac{\s}{\k^{2|\a|}}
  \frac{\G(1+|\a|)}{\G(1-|\a|)}\frac{d}{d(r^{|2\a|})}[r^{|\a|}R_0(r)]\right\}
  =0\,,
\ee
with the corresponding solution given by
\be \label{sol}
R_0(r) = \mbox{(const.)}\left[
         J_{|\a|}(kr) + \s \left(\frac{k}{\k}\right)^{2|\a|}J_{-|\a|}(kr)
         \right]\,,
\ee
where $\s=\pm1$ and $\k$ is a scale introduced by the boundary condition.
If $\k\ra\i$ with $\s=+1$ we recover the hard-core boundary condition 
$\ps(0)=0$.\footnote{If $\s=-1$, formally we obtain $\ps(0)=0$ when $\k\ra\i$
but it does not correspond to the usual hard-core limit. Actually, the system
is not bounded from below because it has a bound state with negative infinite
energy as seen in (\ref{bound}) below \cite{MT}.}
In addition
to the solution (\ref{sol}), if $\s=-1$, there is a bound state satisfying 
(\ref{bc}),
\be \label{bound}
R_0(r)= \mbox{(const.)}K_{|\a|}(\k r)\,,
\ee
with energy $E_B=-\k^2/m$. (Thus the parameter $\k$ in the boundary condition
appears in the bound state energy.) To proceed, we introduce a
circular boundary at radius $R$ and assume $R_0(R)=0$ as usual \cite{arovas}. 
The raduis $R$ will be taken to be infinity
eventually. Then the allowed energies are 
\bea
E_{n,s}&=&\frac{k_{n,s}^2}{m}\quad (n\neq 0)\,,\no\\
E_{0,s}&=&\frac{\t{k}_{s}^2}{m}\quad (n = 0)\,,
\eea
where $k_{n,s}R$ is the $s$-th zero of $J_{|n+\a|}(kR)=0$ and $\t{k}_sR$ is 
the $s$-th zero of (\ref{sol}), i.e.,
\be \label{zero}
J_{|\a|}(kR) + \s \left(\frac{k}{\k}\right)^{2|\a|}J_{-|\a|}(kR) = 0\,.
\ee
Hence, compared with the hard-core case, only $s$-wave energy spectrum is 
changed.

{}From Eqs. (\ref{bt}) and (\ref{4}), the second virial coefficient can then
be written as
\be \label{virial}
B(T)=B_0(T) - 2\l_T^2\left\{ e^{-\b E_B}\th(-\s) + 
     \lim_{R\ra\i}\sum_{s=0}^\i \left[ 
         e^{-\b\frac{\t k_s^2}{m}}-e^{-\b\frac{k_{0,s}^2}{m}}\right]\right\}\,,
\ee
where $\b=1/k_BT$ and $B_0(T)$ is the second virial coefficient for hard-core
anyons calculated in ref.\ \cite{arovas},
\be
B_0(T) = \l_T^2\left(-\frac14 + |\a| - \frac{\a^2}{2}\right)\,.
\ee
To perform the summation in (\ref{virial}) in the limit $R\ra\i$, we rewrite
(\ref{zero}) using 
  $J_{-|\a|}(x) = \cos{\pi\a} J_{|\a|}(x) - \sin{\pi|\a|} N_{|\a|}(x)$,
\be
\cos\eta(k)J_{|\a|}(kR) - \sin{\eta(k)}N_{|\a|}(kR) = 0\,,
\ee
where $\eta(k)$ is defined as
\be \label{eta}
\tan\eta(k) = \frac{\s\left(\frac{k}{\k}\right)^{2|\a|}\sin\pi|\a|}%
                   {1+\s\left(\frac{k}{\k}\right)^{2|\a|}\cos\pi\a}\,.
\ee
Then, from the asymptotic behavior of Bessel functions \cite{AS},
\bea
J_{|\a|}(x) &=&\sqrt{\frac2{\pi x}}
 \cos\left[ x-\left(|\a|+\frac12\right)\frac\pi2\right]+\cdots\,,\no\\
N_{|\a|}(x) &=&\sqrt{\frac2{\pi x}}
 \sin\left[ x-\left(|\a|+\frac12\right)\frac\pi2\right]+\cdots\,,
\eea
it is easy to see that, for any $\e>0$, there exists an integer $s_0$
independent of $R$ such that 
\be
\left|\cos\left[\t{k}_s R - \left(|\a|+\frac12\right)\frac\pi2 
               + \eta(\t k_s)\right]\right|
 \le \sin\e\,,\quad \mbox{if $s \ge s_0$.}
\ee
In other words, if $s \ge s_0$,
\be  \label{app}
\t k_s R + \eta(\t k_s) = s\pi + \left(|\a|-\frac12\right)\frac\pi2 + O(\e)\,.
\ee
Now we split the summation into two parts
$
\sum_{s=0}^\i = \sum_{s=0}^{s_0-1} + \sum_{s=s_0}^\i\,.
$
The first term will vanish eventually as $R\ra\i$ and so the second term
dominates the whole summation. For $s\ge s_0$, we may approximate 
$\t k_s$'s in
the summation by (\ref{app}) (omitting $O(\e)$ term) and the error of the
approximate sum is easily seen to be $O(\e\cdot R^0)$, 
which vanishes as $\e\ra 0$ and $R\ra\i$. This allows us to
replace $\t k_s$ by a function $\t k(s)$ defined for real $s$ as
\be \label{ks}
\t k(s) R + \eta(\t k(s)) = s\pi + \left(\a - \frac12\right) \frac{\pi}2\,,
\ee
when $s\ge s_0$. [$k(s)$ is defined similarly.] Then by use of 
Euler-Maclaurin formula \cite{AS}, 
the summation may be converted into an integral,
\be 
\sum_{s=0}^\i \left[ e^{-\frac{\b\t k_s^2}{m}} 
             - e^{-\frac{\b k_{0,s}^2}{m}}\right]
 = \int_{s_0}^\i ds\left[ e^{-\frac{\b\t k^2(s)}{m}}
             - e^{-\frac{\b k^2(s)}{m}}\right] 
   + O(\e\cdot R^0) + O(1/R)\,,
\ee
where we did not write explicitly other terms in the Euler-Maclaurin formula
which are $O(1/R)$ at most. From (\ref{ks}), 
$\frac{ds}{d\t k} = \frac1\pi(R+\frac{d\eta}{d\t k})$ and 
$\frac{ds}{d k} = \frac{R}\pi$. Thus we get\footnote{With the second term
$\frac1\pi\eta(0)$, (\ref{integral}) generalizes the usual formula 
in terms of phase shifts \cite{UB} whose
naive application does not produce the correct result in anyon case as 
noted in \cite{CGO}.}
\be \label{integral}
\lim_{R\ra\i}\sum_{s=0}^\i \left[ e^{-\frac{\b\t k_s^2}{m}}
             - e^{-\frac{\b k_s^2}{m}}\right]
 = \frac1\pi\int_0^\i dk \frac{d\eta}{d k}e^{-\frac{\b k^2}{m}}
                         +\frac1\pi \eta(0)\,.
\ee
From (\ref{eta}), $\eta(0)=0$ unless $\k=0$ and so we drop this term from
now on assuming $\k\neq0$\footnote{When $\k=0$, $\eta=\pi|\a|$ and the first
term involving $\frac{d\eta}{dk}$ becomes zero.}. 
$\frac{d\eta}{dk}$ can easily be computed from (\ref{eta}). 
Inserting this into (\ref{integral}), we finally obtain 
\be
B(T) = B_0(T) - 2\l_T^2\left\{ e^\ve\th(-\s)
       +\frac{|\a|\s}{\pi}\sin{\pi|\a|}
        \int_0^\i \frac{dt e^{-\ve t} t^{|\a|-1}}%
                       {1+2\s\cos{\pi\a}t^{|\a|}+t^{2|\a|}} \right\}\,,
\ee
where $\ve=\b\k^2/m$. If we put $\s=-1$, this result is equivalent to the
formulas in refs.\ \cite{moroz,GMS}.

The above integration cannot be performed analytically in general. In case of
semions, $|\a|=1/2$, it is reduced to the error integral,
\be 
B(|\a|=1/2,T) = B_0(|\a|=1/2,T) - \l_T^2 e^\ve [1-\s\erf(\sqrt\ve)]\,,
\ee
where $\erf(x) = \frac{2}{\sqrt{\pi}}\int_0^x e^{-t^2} dt$\,.
Compared with $B_0(|\a|=1/2,T)=\l_T^2/8$, the second term 
becomes dominant for most of $\ve$ except near the hard-core limit 
$\ve\ra\i$ with $\s=+1$.

For general value of $\a$, one can obtain various limiting behaviors by
examining the integral. First, for $\ve \gg 1$, we find the asymptotic 
behavior 
\be \label{egg1}
B(T)=B_0(T) -2\l_T^2 \left[ e^\ve \th(-\s)
  +\frac{\s}{\pi}\frac{\G(|\a|+1)}{\ve^{|\a|}}\sin{\pi|\a|}+\cdots \right]\,.
\ee
We see that, for $\s=+1$, the hard-core result is recovered but, if $\s=-1$,
the bound state contribution dominates the virial coefficient. 
Because of the nature of the virial expansion, (\ref{egg1}) is physically 
meaningful only for parameters in the region
$\k^2/m \gg k_BT \gg \r/m$ where $\r$ is the particle number density.
For $\ve\ll 1$, which corresponds to high-temperature limit, we find after
some calculation
\bea
B(T)&=&B_0(T) - 2\l_T^2 |\a| ( 1- \s \ve^{|\a|} + \cdots ) \,,\no\\
    &=& \l_T^2\left( -\frac14 - |\a| - \frac{\a^2}{2} \right)
        + 2 \s \l_T^2 |\a| \ve^{|\a|} + \cdots \,,
\eea
which differs from the hard-core result by $-2\l_T^2 |\a|$ for $\ve = 0$.

One may also consider the behavior of $B(T)$ near the bosonic 
($\a=0$) and the fermionic ($|\a|=1$) points. After some calculation, we
obtain\footnote{When $\s=-1$ and $\a=0$ 
this result gives the second virial coefficient of bosons with the 
self-adjoint extension discussed in \cite{BL,GMS}.},
for $\a\approx0$ and for nonzero finite $\ve$, 
\bea
B(T)&=&B_0(T) - 2\l_T^2\n(\ve)\th(-\s) - |\a|\l_T^2+ O(\a^2)\no\\
    &=&-\left[ \frac14 + 2\n(\ve)\th(-\s) \right]\l_T^2 + O(\a^2) \,,
\eea
where $\n(\ve)$ is the Neumann function defined by \cite{GR}
\be
\n(\ve) = \int_0^\i\frac{\ve^t dt}{\G(t+1)}\,.
\ee
Therefore the linear terms in $|\a|$ are cancelled out, i.e., $B(T)$ is a
smooth function of $\a$ near $\a=0$ and the cusp at the
bosonic point existing for hard-core anyons disappears. On the other hand,
near the fermionic point, we find
\bea
B(T)&=&B_0(T) - 2\l_T^2 e^{-\s\ve} + f_\s(\ve) (1-|\a|) + \cdots\no\\
    &=&\l_T^2 \left[ -\frac12 (1-|\a|)^2 +\frac14 - 2 e^{-\s\ve}\right] 
        + f_\s(\ve) (1-|\a|) + \cdots \,,
\eea
where $f_\s(\ve)$ ($\s=\pm1$) are some complicated functions of $\ve$ 
involving exponential integrals whose explicit forms are not much 
illuminating. Thus we
see that in general a cusp exists at the fermionic point $|\a|=1$ for anyons
without hard core. 

The physical reason for this change of the cusp position may be the following.
For the hard-core boundary condition, anyons behave more like fermions in the
sense that they cannot overlap and the lack of hard core for bosons is
responsible for the cusp \cite{LF}. In contrast, for other boundary
conditions, anyons can overlap like bosons and this generates a cusp at the
fermionic point where particles cannot overlap.

We plot these results in Figure 1. The deviation of the
virial coefficient from hard-core result is quite significant for
reasonable values of $\ve = \b\frac{\k^2}{m}$. Also, Figure 1 clearly shows
cusps at $|\a|=1$ and smoothness of curves at $\a=0$.
     
One may be able to generalize this calculation to a system of multi-species
of anyons obeying mutual statistics \cite{mutual}. 
This system is known to be related with multi-layered
quantum Hall effect. In this case self-adjoint extension
is allowed in $p$-wave part as well as in $s$-wave part and it introduces
another scale parameter. Field theoretically it would correspond to
introducing a nonrenormalizable contact interaction with derivatives of delta
functions. It may be interesting to see the effect of these scale parameters
on various physical properties of the system. Also, the analysis of this paper
may be able to be generalized to a system of non-Abelian Chern-Simons 
particles for which the second virial coefficient was recently calculated 
in the hard-core case\cite{tlee}.

We would like to thank Prof. C.\ Lee for bringing my attention to this topic 
and for continuous encouragement, and S.\ J.\ Kim for useful
discussions. We also thank A.\ Moroz and R.\ Soldati for informing us of their
works. This work was supported by the Korea Science 
and Engineering Foundation through the SRC program.

\begin{figure}
\vspace{1cm}
\epsfxsize=14cm
\centerline{\epsffile{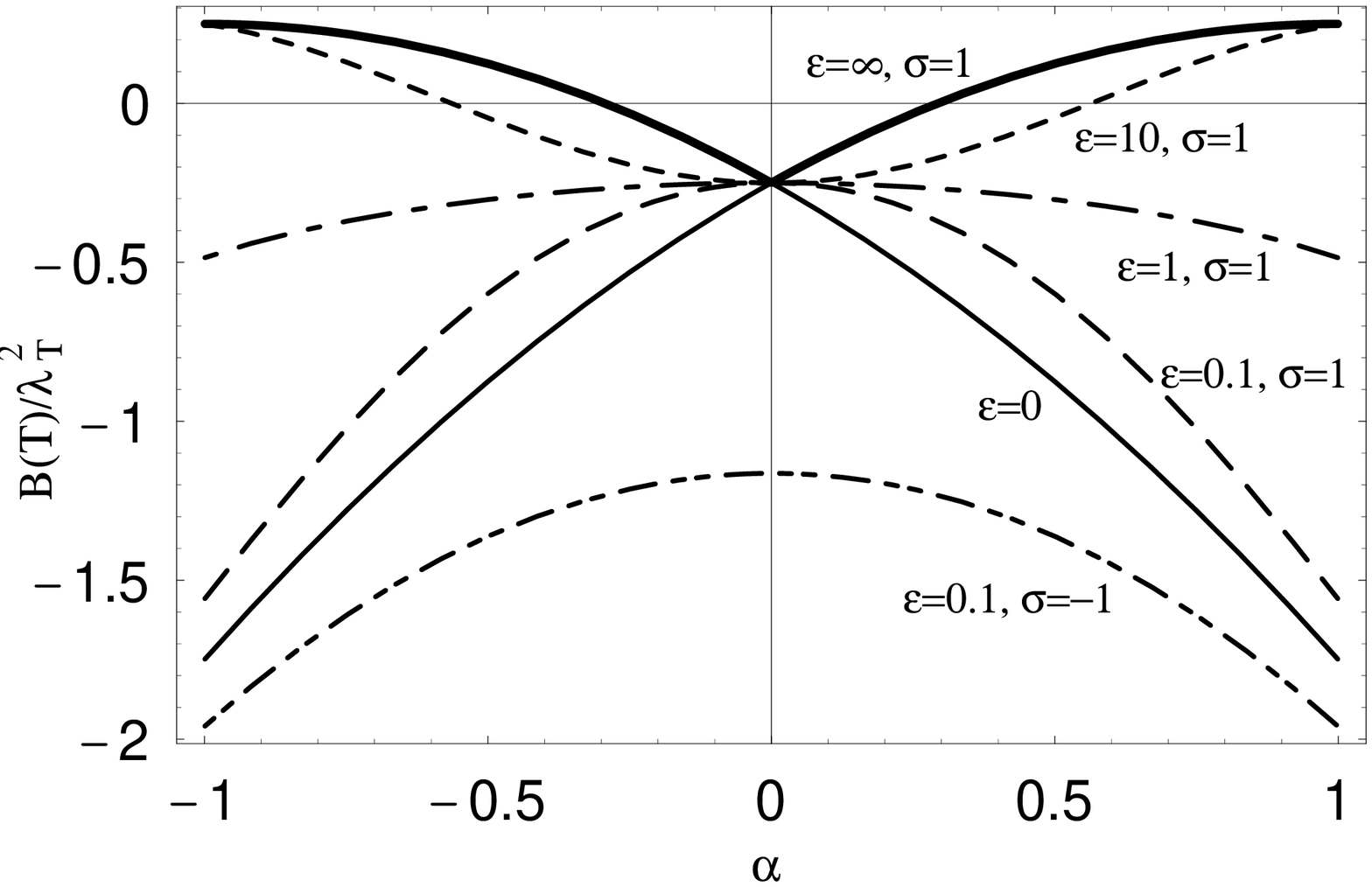}}
\vspace{1cm}
\caption{The second virial coefficient of anyons without hard core as
functions of the statistical parameter $\a$
for various values of $\ve=\b\k^2/m$.
$\ve=\i$ with $\s=1$ corresponds to hard-core limit 
and $\ve=0$, high-temperature limit.}
\end{figure}

\begin{references}
\bibitem{anyon} J.\ M.\ Leinaas and J.\ Myrhein, Nuovo Cimento B {\bf 37}
(1977) 1; G.\ A.\ Goldin, R.\ Menikoff, and D.\ H.\ Sharp, \jmp{21} (1980) 21;
F.\ Wilczek, \prlt{48} (1982) 1144; for a review, see F.\ Wilczek, {\it
Fractional Statistics and Anyonic Superconductivity} (World Scientific, 1990).
\bibitem{SA} S.\ Albeverio, F.\ Gesztesy, R.\ H{\o}egh-Krohn, and H.\ Holden,
{\it Solvable Models in Quantum Mechanics} (Springer-Veerlag, Berlin, 1988).
\bibitem{AB} Y.\ Aharonov and D.\ Bohm, \pr{115} (1959) 485.
\bibitem{hagen} C.\ R.\ Hagen, \pr{D31} (1985) 848.
\bibitem{JP} R.\ Jackiw and S.\ Y.\ Pi, \pr{D42} (1990) 3500.
\bibitem{BL} R.\ Jackiw, in {\it M.\ A.\ B\'eg Memorial Volume}, edited by
A.\ Ali and P.\ Hoodbhoy  (World Scientific, Singapore, 1991); O.\ Bergman and
G.\ Lozano, \anp{229} (1994) 416.
\bibitem{AM} G.\ Amelino-Camelia, \plt{326B} (1994) 282; C.\ Manuel and R.\
Tarrach, \plt{328B} (1994) 113; S.\ Ouvry, \pr{D50} (1994) 5296;
G.\ Amelino-Camelia, \pr{D51} (1995) 2000.
\bibitem{AD} G.\ Amelino-Camelia and D.\ Bak, \plt{343B} (1995) 231.
\bibitem{KL} S.\ J.\ Kim \plt{343B} (1995) 244; S.\ J.\ Kim and C.\ Lee
       \pr{D55} (1997) 2227.
\bibitem{BoS} M.\ Bourdeau and R.\ D.\ Sorkin, \pr{D45} (1992) 687.
\bibitem{BB} D.\ Bak and O.\ Bergman, \pr{D51} (1995) 1994.
\bibitem{arovas} D.\ P.\ Arovas, R.\ Schrieffer, F.\ Wilczek and A.\ Zee, 
  \np{B251} (1985) 117.
\bibitem{CGO} A.\ Comtet, Y.Georgelin and S.\ Ouvry, \jp{A22} (1989) 3917.
\bibitem{LF} D.\ Loss and Y.\ Fu, \prlt{67} (1991) 294.
\bibitem{HK} See also T.\ Blum, C.\ R.\ Hagen and S.\ Ramaswamy, \prlt{64}
(1990) 709; B.\ Kahng and K.\ Park, \pr{B45} (1991) 8158.
\bibitem{moroz} A.\ Moroz, \pr{A53} (1996) 669.
\bibitem{GMS} P.\ Giacconi, F.\ Maltoni and R.\ Soldati, \pr{B53} (1996)
10065.
\bibitem{huang} See for example K.\ Huang, {\it Statistical Mechanics} (John
Wiley and Sones, 1963).
\bibitem{MT} C.\ Manuel and R.\ Tarrach, \plt{268B} (1991) 222. 
\bibitem{AS} M.\ Abramowitz and I.\ Stegun, {\it Handbook of  Mathematical
Functions} (Dover, 1972).
\bibitem{UB} E.\ Bethe and G.\ E.\ Ulenbeck, Physica {\bf 4} (1937) 915.
\bibitem{GR} I.\ S.\ Gradstein and I.\ M.\ Ryzik, {\it Tables of integrals,
series and products} (Academic Press, 1980).
\bibitem{mutual} X.\ G.\ Wen and A.\ Zee, \np{B15} (1990) 135; \pr{B44} (1991)
274; B.\ Blok and X.\ G.\ Wen, \pr{B43} (1991) 8337; 
F.\ Wilczek, \prlt{69} (1992) 132; C.\ R.\ Hagen \prlt{68} (1992) 3821; 
C.\ Kim, C.\ Lee, P.\ Ko, B.-H.\ Lee, and H.\ Min, \pr{D48} (1993)1821;
D.\ Wesolowski, Y.\ Hosotani, and C.-L.\ Ho, \ijmp{A9} (1994) 969.
\bibitem{tlee} T.\ Lee, \prlt{74} (1995) 4967.
\end{references}
\end{document}